\documentclass[12pt]{iopart}
\usepackage{epsfig}

\begin{document}
\title{Thermal photons from fluctuating initial conditions}

\author{\underline{Rupa Chatterjee}$^1$, Hannu Holopainen$^{1,2}$, Thorsten Renk$^{1,2}$, and Kari J. Eskola$^{1,2}$}

\address{$^{1}$Department of Physics, P. O. Box 35, FI-40014 University of Jyv\"askyl\"a, Finland}
\address{$^2$Helsinki Institute of Physics, P.O.Box 64, FI-00014 University of Helsinki, Finland}

\ead{rupa.r.chatterjee@jyu.fi}

\begin{abstract}
Event-by-event fluctuations of initial QCD-matter density produced in heavy-ion collisions at RHIC enhance the production of thermal photons significantly in the region $2 \le p_T \le 4$ GeV/$c$ compared to a smooth initial-state averaged profile in the ideal hydrodynamic calculation. This enhancement is a an early time effect due to the presence of hotspots or over-dense regions in the fluctuating initial state. The effect of fluctuations is found to be stronger in peripheral than in central collisions. 
\end{abstract}



\section{Introduction}

Photons have long been considered as one of the most powerful probes to characterize the initial state of the system produced in ultra-relativistic heavy-ion collisions~\cite{phot}. This is complementary to the $p_T$ spectra of bulk hadrons which reflect the late conditions close to freeze-out. The thermal emission of photons shows a strong temperature dependence, high $p_T$ photons are mostly emitted from the hot and dense early stage of the system evolution when the hydrodynamic flow is weak, whereas low $p_T$ photons are emitted from the flow boosted relatively cold later part of the system. Recent studies have shown that event-by-event QCD-matter density fluctuations in the initial conditions (IC) are needed (in addition to a correct reference plane definition) for reproducing the measured charged-particle elliptic flow even for most central collisions, which was underestimated by all earlier hydrodynamic calculations using smooth IC~\cite{hannu}. 

We show that fluctuations in the initial state enhance the production of thermal photons significantly for  $p_T > 1$ GeV/$c$ compared to a smooth IC. This makes the thermal photons in that $p_T$ range especially suitable to probe the hotspots in the IC~\cite{fluctuation}.

\section{Event-by-event hydrodynamics and thermal photons}
We use the event-by-event hydrodynamical model developed in~\cite{hannu} to calculate the production of thermal photons from fluctuating as well as from smooth IC. A Monte Carlo Glauber model with the standard two-parameter Woods-Saxon nuclear density profile is used to create the initial state nucleon configurations, and to compute the number of wounded nucleons. The initial density profile is taken to be proportional to the number of wounded nucleons (WN), where  entropy density $s$  is distributed in the $(x,y)$ plane around the wounded nucleons using a 2D Gaussian smearing,
\begin{equation}
  s(x,y) = \frac{K}{2 \pi \sigma^2} \sum_{i=1}^{\ N_{\rm WN}} \exp \Big( -\frac{(x-x_i)^2+(y-y_i)^2}{2 \sigma^2} \Big).
 \label{eq:eps}
\end{equation}
Here $K$ is a fixed overall normalization constant and $\sigma$ is a free  parameter which determines the size of the fluctuations. 
The initial formation time of the plasma  is taken as $\tau_0=$ 0.17 fm/$c$, motivated by the EKRT minijet saturation model~\cite{ekrt}. We choose a default value for the size parameter as $\sigma=$ 0.4 fm~\cite{hannu}. The equation of state which shows a sharp cross-over transition from the plasma phase to the hadronic phase is from~\cite{eos}. 
The temperature at freeze-out is taken as 160 MeV (see~\cite{hannu} and~\cite{fluctuation} for more details). 
This model has been applied successfully to reproduce both the measured centrality dependence as well as the $p_T$ dependence of charged particle elliptic flow up to $p_T \sim 2$ GeV/$c$, and the pion $p_T$ spectra up to $\sim 3$~GeV/$c$~\cite{hannu}.

We use the complete leading order rates $R=EdN/d^3p d^4x$ of~\cite{amy} for the emission from plasma and the parametrization given in~\cite{trg} for the hadronic matter emission which at present can be considered as the state of the art for photon production. The switching from the plasma rates to hadronic rates is done at a temperature 170 MeV.
The total thermal emission from the produced QCD matter is obtained by integrating the emission rates over the space-time history of the fireball as  
\begin{equation}
\label{eq1}
  E\, dN/d^3p = \int d^4x\, R\Big(E^*(x),T(x)\Big),
\end{equation}
with $T(x)$ as the local temperature and $E^*(x)=p^{\mu}u_{\mu}(x)$, where $p^\mu$ is the four-momentum of the photon and  $u^\mu$ is the local four-velocity of the flow field.
\begin{figure}
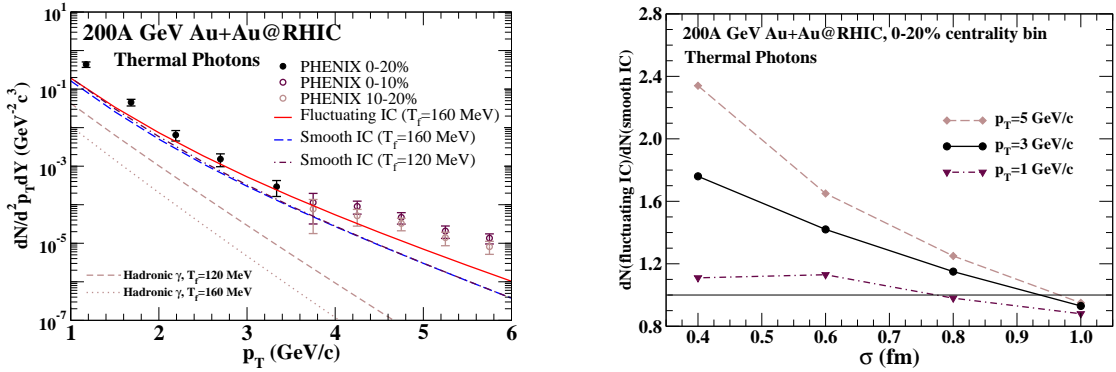

\epsfig{file=phot.eps, width=6.8cm,clip=true} \hspace{1.2cm}\epsfig{file=sigma.eps, width=6.6cm,clip=true}
\caption{(From \cite{fluctuation}. Color online) [Left] Thermal photons from a smooth (long dashed blue line) and fluctuating (solid red line) IC along with PHENIX direct photon data \cite{phenix1,phenix2} for 0--20\% centrality. [Right] $\sigma$ dependence of the results.}
\label{fig1}
\end{figure}

\begin{figure}
\centerline{\includegraphics*[width=12.0cm]{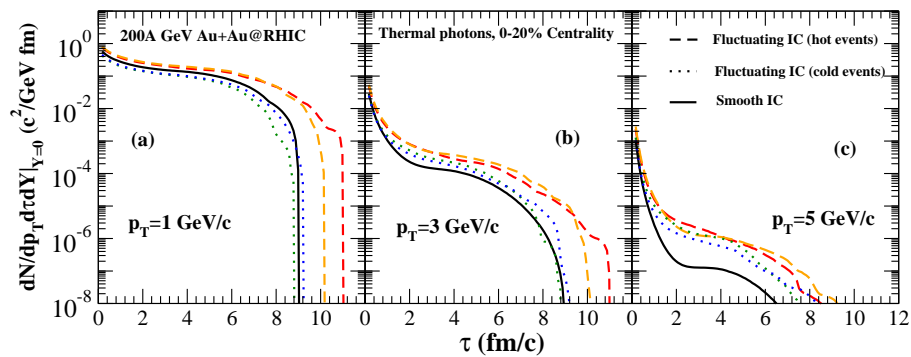}}
\caption{(From~\cite{fluctuation}. Color online) Time evolution of thermal photon yield for $p_T$ values of (a) 1, (b) 3, and (c) 5 GeV/$c$. Results are compared with  an initial state averaged event and four different random events from the fluctuating IC.}
\label{fig2}
\end{figure}
\section{Results and discussions}
Figure~\ref{fig1} (left panel) shows our results for thermal photon spectra from a smooth and from fluctuating IC for 200A GeV Au+Au collisions at RHIC along with PHENIX data~\cite{phenix1,phenix2} for the 0-20 \% centrality bin. 
The result from the fluctuating IC scenario is obtained by averaging photon spectra from a sufficiently large number of random events. The smooth profile is obtained by taking an average of 1000 initial profiles which is enough to remove essentially all the fluctuations~\cite{fluctuation}. We see that for both the smooth and the fluctuating IC, the spectra are dominated by radiation from the QGP phase in the entire $p_T$ range shown in the figure. We further notice that even with a much smaller freeze-out temperature ($\sim$~120 MeV) we get significant contributions only below   $p_T \sim$ 1.5 GeV/$c$ as shown by the brown dashed dotted line in Figure~\ref{fig1}. 

The slope of the photon $p_T$ spectrum from the fluctuating IC is about 10\% flatter compared to the slope from the smooth IC in the region $2 \le p_T \le 4$ GeV/$c$. For $p_T < 2$ GeV/$c$, with the fluctuating IC we obtain 20-40\% more photons than with the smooth profile, whereas for $p_T > 2$ GeV/$c$ the two results differ almost by a factor of 2. 
We know that the thermal emission of photons is exponential in temperature and linear in radiating volume. As a result, the hotspots in the fluctuating IC produce more high $p_T$ photons than the smooth profile and the difference between the two IC increases towards higher values of transverse momentum. We find that the photon results from the fluctuating IC show a better agreement with the PHENIX  experimental data for $2 \le p_T \le 4$ GeV/$c$ leaving enough space for the other sources of direct photons besides the thermal contribution. 

The thermal photon production from the fluctuating IC is found to be quite sensitive to the value of the fluctuation size parameter $\sigma$, where the enhancement in the production compared to a smooth IC is found to be maximal towards smaller values of $\sigma$ and also for larger values of $p_T$ (see right panel of Figure~\ref{fig1}). 

The time evolution of thermal photon yield for different values of $p_T$ is shown in Figure~\ref{fig2} which clearly explains the dynamics leading to the results presented in Figure~\ref{fig1}. We see that most of the high $p_T$ photons are emitted from very early stage of the system expansion and their production drops by a few order of magnitude within a couple of fm/$c$. We further notice that the cold events (or events having less than average entropy) produce more high $p_T$ photons than the smooth IC for $p_T \ge 3$ GeV/$c$. 

\begin{figure}[t]
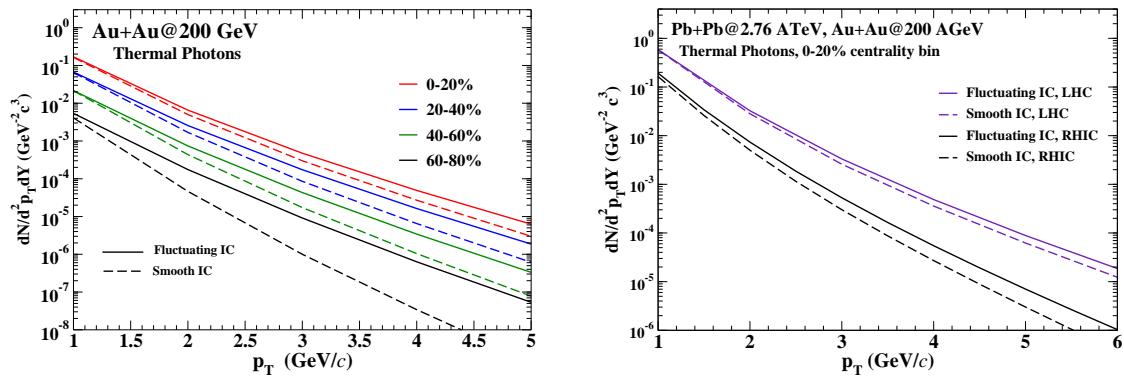

\epsfig{file=rhic_centrality.eps, width=7.0cm,clip=true}\hspace{.8 cm}\epsfig{file=rhic_lhc.eps, width=7.0cm,clip=true}
\caption{(Color online) [Left] Centrality dependence of the thermal photon spectra from smooth and fluctuating IC. [Right] Results at RHIC and LHC are compared.}
\label{fig3}
\end{figure}

The centrality dependence of the results is shown in Figure~\ref{fig3} (left panel) where the difference between the smooth and fluctuating IC is found to increase from central to peripheral collisions. 
The size of the produced system becomes smaller towards peripheral collisions and the effect of the hot spots in the fluctuating IC  becomes more pronounced, leading to the shown result.
We have also noticed that the effect of initial state fluctuations in the excess production of high $p_T$ photons is weaker at LHC than at RHIC~\cite{chre} (see right panel of Figure~\ref{fig3}). 

In conclusion, we see that fluctuations in the initial density distribution enhance significantly the production of thermal photons compared to a smooth initial state averaged profile in ideal hydrodynamic calculation. This is an early time effect solely due to the presence of hotspots in the fluctuating IC and this effect is found to be stronger towards peripheral collisions. This calls for a more detailed study of the effects of fluctuations for lower beam energies and for smaller size systems. In addition to that, estimation of photon elliptic flow from the fluctuating IC will be very interesting. 

We are financially supported by the Academy of Finland (projects  130472 and 133005), by the national Graduate School of Particle and Nuclear Physics, and by a Magnus Ehrnrooth Foundation travel grant. We acknowledge CSC-IT Center for Science in Espoo, Finland, for computational resources.

\end{document}